\documentclass[aps,superscriptaddress,twocolumn,longbibliography,pra]{revtex4-1}
\usepackage{bm,bbold} 
\usepackage{graphicx} 
\usepackage{amsmath}
\usepackage{comment}
\usepackage{amsthm}
\usepackage{amssymb} 
\usepackage{epstopdf} 
\usepackage{mathtools}
\usepackage{amsfonts}
\usepackage{epsfig}
\usepackage{color} 
\usepackage{subfigure}
\usepackage[english]{babel}
\usepackage{mathtools}
\usepackage{hyperref}
\hypersetup{
    colorlinks=true,       
    linkcolor=cyan,          
    citecolor=magenta,        
    filecolor=magenta,      
    urlcolor=cyan,           
    runcolor=cyan
}

\newcommand{\bra}[1]{\langle #1 |}
\newcommand{\ket}[1]{| #1 \rangle}

\newcommand {\be}{\begin{equation}}
\newcommand {\ee}{\end{equation}}

\newcommand{\ba}{\begin{eqnarray}}
\newcommand{\ea}{\end{eqnarray}}

\newcommand{\ignore}[1]{}

\usepackage[margin=1in,nofoot]{geometry}

\newcommand{\e}{{{e}}}
\newcommand{\rmd}{{\text d}}

\newcommand{\beq}{\begin{equation}}
\newcommand{\eeq}{\end{equation}}
\newcommand{\beqnn}{\begin{equation*}}
\newcommand{\eeqnn}{\end{equation*}}
\newcommand{\bea}{\begin{eqnarray}}
\newcommand{\eea}{\end{eqnarray}}
\newcommand{\beann}{\begin{eqnarray*}}
\newcommand{\eeann}{\end{eqnarray*}}
\newcommand{\bes} {\begin{subequations}}
\newcommand{\ees} {\end{subequations}}

\newcommand {\nn}{\nonumber}

\begin{document}

\title{An integral-free representation of the Dyson series using divided differences}
\author{Amir Kalev}
\affiliation{Information Sciences Institute, University of Southern California, Arlington, VA 22203, USA}
\author{Itay Hen}
\affiliation{Information Sciences Institute, University of Southern California, Marina del Rey, CA 90292, USA}
\affiliation{Department of Physics and Astronomy, and Center for Quantum Information Science \& Technology,University of Southern California, Los Angeles, California 90089, USA}

\begin{abstract}
\noindent The Dyson series is an infinite sum of multi-dimensional time-ordered integrals, which serves as a formal representation of the quantum time-evolution operator in the interaction-picture. Using the mathematical tool of divided differences, we introduce an alternative representation for the series that is entirely free from both time ordering and integrals. In this new formalism, the Dyson expansion is given as a sum of efficiently-computable divided differences of the exponential function, considerably simplifying the calculation of the Dyson expansion terms, while also allowing for time-dependent perturbation calculations to be performed directly in the  Schr{\"o}dinger-picture. We showcase the utility of this novel representation by studying a number of use cases. We also discuss several immediate applications. 
\end{abstract}

\maketitle
\section{Introduction}
The Dyson series~\cite{Dyson49} is one of the fundamental results of quantum scattering theory. It is a perturbative expansion of the quantum time-evolution operator in the interaction-picture, wherein each summand is formally represented as a multi-dimensional integral over a time-ordered product of the interaction-picture Hamiltonian at different points in time. As such, the series serves as a pivotal tool for studying transition properties of time-dependent quantum many-body systems~\cite{Fetter}. The Dyson series also has a close relation to Feynman diagrams in quantum field theory~\cite{Weinberg} and to the Magnus expansion in the theory of first-order homogeneous linear differential equation for linear operators~\cite{Blanes}. 

Despite its fundamental role in quantum theory, one of the key challenges in using the Dyson series in practical applications remains the evaluation of the multi-dimensional integrals over products of time-ordered operators, making the calculation of the terms in the series an exceedingly complicated task~\cite{DysonTree}. In this paper, we derive an analytical, closed-form expression for the summands of the Dyson series, by explicitly evaluating the Dyson integrals.  We accomplish this by utilizing the machinery of `divided differences' -- a mathematical tool normally used for computing tables of logarithms and trigonometric functions and for calculating the coefficients in the interpolation polynomial in the Newton form~\cite{Thomson1933,dd:67,deboor:05,ODE,ODE2,pmr,2020arXiv200602539K}. 

The main technical contribution of this work is an alternative, yet equivalent, formulation of the Dyson series wherein the summands are given in terms of efficiently computable divided differences of the exponential function. We argue that our novel representation, which is devoid of integrals and time ordering, makes it a very useful tool in the study of perturbation effects in many-body quantum systems --- a fundamental branch of quantum physics. 

In particular, this representation allows us to write an explicit, integral-free time-dependent perturbation expansion for the time-evolution operator in the Schr{\"o}dinger-picture enabling us to carry perturbation calculations without the need to switch to the interaction-picture. To showcase the utility of our formulation, we work out a number of examples, for which we explicitly calculate the first few terms of the series. 
We begin by a brief introduction of divided differences, followed by the derivation of the integral-free representation of the Dyson series.  

\section{Divided differences --  a brief overview \label{sec:dd}}
The divided differences~\cite{dd:67,deboor:05}, which will be a major ingredient in the derivation detailed in what follows, is a recursive division process.  The divided differences of a function $f(\cdot)$ is defined as
\beq\label{eq:divideddifference2}
f[x_0,\ldots,x_q] \equiv \sum_{j=0}^{q} \frac{f(x_j)}{\prod_{k \neq j}(x_j-x_k)}
\eeq
with respect to the list of real-valued input variables $[x_0,\ldots,x_q]$. The above expression is ill-defined if some of the inputs have repeated values, in which case one must resort to the use of limits. For instance, in the case where $x_0=x_1=\ldots=x_q=x$, the definition of divided differences reduces to: 
\beq
f[x_0,\ldots,x_q] = \frac{f^{(q)}(x)}{q!} \,,
\eeq 
where $f^{(n)}(\cdot)$ stands for the $n$-th derivative of $f(\cdot)$.
Divided differences can alternatively be defined via the recursion relations
\bea\label{eq:ddr}
f[x_i,\ldots,x_{i+j}] = \frac{f[x_{i+1},\ldots , x_{i+j}] - f[x_i,\ldots , x_{i+j-1}]}{x_{i+j}-x_i} \,,\nonumber\\
\eea 
with $i\in\{0,\ldots,q-j\},\ j\in\{1,\ldots,q\}$ and the initial conditions
\beq\label{eq:divideddifference3}
f[x_i] = f(x_{i}), \qquad i \in \{ 0,\ldots,q \}  \quad \forall i \,.
\eeq
As is evident from the above, the divided differences of any given function $f(\cdot)$ with $q$ inputs can be calculated with  $q(q-1)/2$ basic operations.

{Divided differences obey the following properties:
(i) Linearity: for any two functions $f(\cdot)$ and $g(\cdot)$ and a number $\lambda$, the following holds. \hbox{$(f+g)[x_0,\ldots,x_q] =f[x_0,\ldots,x_q]+g[x_0,\ldots,x_q]$} and \hbox{$(\lambda \cdot f)[x_0,\ldots,x_q] =\lambda \cdot f[x_0,\ldots,x_q]$}. 
(ii) Divided differences obey the Leibniz rule \hbox{$(f \cdot g)[x_0,\ldots,x_q] =\sum_{k=0}^q f[x_0,\ldots,x_k] \cdot g[x_k,\ldots,x_q]$}. 
(iii) Invariance under permutation of inputs, namely, ${\displaystyle f[x_{0},\dots ,x_{q}]=f[x_{\sigma (0)},\dots ,x_{\sigma (q)}]}$ where ${\displaystyle \sigma :\{0,\dots ,q\}\to \{0,\dots ,q\}}$ is a permutation. 
(iv) Any real-valued function $f(\cdot)$ obeys the mean-value theorem $f[x_0,\ldots,x_q] = f^{(q)}(\xi)/q!$ where $f^{(q)}(\cdot)$ stands for the $q$-th derivative of $f(\cdot)$ and $\xi$ is a number in the interval $[\min[x_0,\ldots,x_q],\max[x_0,\ldots,x_q]]$.
}

In addition, a function of divided differences can be defined in terms of its Taylor expansion. For the special case of  
$f(x)=\e^{-i t x}$, which will be of special importance in this study, we have
\beq
\e^{-i t [x_0,\ldots,x_q]} = \sum_{n=0}^{\infty} \frac{(-i t)^n [x_0,\ldots,x_q]^n}{n!} \ . 
\eeq 
Moreover, it is easy to verify that
\beq  \label{eq:ts}
[x_0,\ldots,x_q]^{q+n} = \Bigg\{ 
\begin{tabular}{ l c l }
   $0$ & \phantom{$0$} & $n<0$ \\
  $1$ & \phantom{$0$} &  $n=0$ \\
  $\sum_{\sum k_j = n} \prod _{j=0}^{q} x_j^{k_j}$& \phantom{$0$} &  $n>0$ \\
\end{tabular}
 \,.
\eeq
One may therefore write:
\bea
\e^{-i t[x_0,\ldots,x_q]} &=& \sum_{n=0}^{\infty} \frac{(-i t)^n [x_0,\ldots,x_q]^n}{n!} \nonumber\\
&=&
\sum_{n=q}^{\infty} \frac{(-i t)^n [x_0,\ldots,x_q]^n}{n!} \nonumber\\
&=&
\sum_{n=0}^{\infty} \frac{(-i t)^{q+n} [x_0,\ldots,x_q]^{q+n}}{(q+n)!}\nonumber \,.
\eea

\section{Divided differences formulation of the time-evolution operator\label{sec:off}}

Usually, the go-to approach to tackling the dynamics of time-dependent systems is through the use of (time-dependent) perturbation theory in the interaction-picture~\cite{Messiah:vol2,Sakurai:book} where the Hamiltonian is written as
\beq\label{eq:ham}
H(t) = H_0 + V(t)\, ,
 \eeq
 namely as a sum of a static (or `free') Hamiltonian $H_0$ (that is assumed to have a known spectrum) and an additional time-dependent Hamiltonian, $V(t)$, that is treated as a perturbation to $H_0$. The time-evolution operator in the interaction-picture is given by \hbox{$U_I(t) = \mathcal{T} \exp[{-i \int_0^t H_I(t) \rmd t}]$}, a shorthand for the Dyson series 
\begin{align}
U_I(t){=}{\sum_{q=0}^{\infty}} {(-i)^q}\!\!{\int_0^{t}} \!\!\!{\rmd t_q} {\cdots} {\int_0^{t_2}} \!\!\! { \rmd t_1} H_I(t_q) \cdots H_I(t_1)\label{eq:dyson},
\end{align}
where $H_I(t)= \e^{i H_0 t} V(t)\e^{-i H_0 t}$, and hereafter we use the $q=0$ term to symbolize the identity operator.

The operator
$U_I(t)$ evolves the interaction-picture wave-function $|\psi_I(t)\rangle$ which is related to the Schr{\"o}dinger-picture wave-function via
\hbox{$|\psi_I(t)\rangle= \e^{i H_0 t}|\psi(t)\rangle$} (in our units, $\hbar=1$). Similarly, the Schr{\"o}dinger-picture time-evolution operator $U(t)$ is related to the interaction-picture operator via $U(t)= \e^{-i H_0 t} U_I(t)$. 
In what follows we present an equivalent form for the Dyson series, Eq.~\eqref{eq:dyson}, by systematically evaluating the integrals in the sum, writing $V(t)$ as a sum of exponentials in $t$.

\subsection{Generalized permutation operator representation of the perturbation Hamiltonian}
We begin by denoting the eigenstates and eigenenergies of  the free Hamiltonian $H_0$ by $\mathcal{B}=\{|z\rangle\}$ and $\mathcal{E}=\{E_z\}$, respectively, such that \hbox{$H_0 | z \rangle = E_z| z \rangle$}. (For simplicity we assume a discrete countable set of eigenstates and eigenenergies). We will refer to $\mathcal{B}$ as the `computational basis'. Next, we write the perturbation Hamiltonian $V(t)$ as a sum of {generalized permutation operators} $\Pi_i$~\cite{gpm}: 
\beq \label{eq:basic}
V(t) =\sum_{i=0}^M \Pi_i(t)=\sum_{i=0}^M D_i(t) P_i \,,
\eeq
where every generalized permutation operator is further expressed as a product of a (time dependent) diagonal (in the computational basis) operator $D_i$ and a bona-fide permutation operator $P_i$. Specifically, the action of $D_i$ and $P_i$ on a computational basis states is given by \hbox{$D_i | z \rangle = d_i(z)| z \rangle$}, where $d_i(z)$ is in general a complex number, and  \hbox{$P_i| z \rangle=| z'\rangle$} for some $\ket{z'}\in\mathcal{B}$ depending on $i$ and $z$. The $i=0$ permutation operator will be reserved to the identity operator, that is, $P_0 =\mathbb{1}$. Equipped with these notations, the action of a  generalized permutation operator  $\Pi_i$ on a basis state $\ket{z}$ is given by \hbox{$D_i P_i | z \rangle = d_i(z')| z' \rangle$}, where $z'$ depends on both the state $z$ and the operator index $i$. We note that any Hamiltonian can be readily cast in the above form~\cite{pmr}. 

At this point, we write each diagonal operator, $D_i(t)$, in Eq.~\eqref{eq:basic}  as an exponential sum in $t$, that is, 
\beq
D_i(t)=\sum_{k=1}^{K_i} \e^{i \Lambda_i^{(k)} t} D_i^{(k)}  \,,
\eeq
where both $\Lambda_i^{(k)}$ and $D_i^{(k)}$ are (generally complex-valued) diagonal matrices and $K_i$ denotes the number of terms in the decomposition of $D_i$. (For more details as to how to carry out this decomposition efficiently, see Refs.~\cite{expSum1,expSum2,doi:10.1063/1.2873123}.) Thus, $V(t)$ can be written as 
\beq\label{eq:hint2}
V(t) =  \sum_{i=0}^M \sum_{k=1}^{K}  \e^{i \Lambda_i^{(k)} t} D_i^{(k)} P_i
\eeq
where, for simplicity, hereafter we fix $K_i=K$ $\forall i$, though this assumption can be easily removed. With this expansion of $V(t)$, we can write the interaction-picture Hamiltonian $H_I(t)$ as
\begin{align}\label{eq:hint3}
H_I(t)=\sum_{i=1}^M \sum_{k=1}^{K}  \e^{i H_0^{(i,k)} t} D_i^{(k)} P_i \e^{-i H_0 t} \,,
\end{align}
where we have defined $H_0^{(i,k)}\equiv H_0+\Lambda_i^{(k)}$. 
Using this form of $H_I(t)$ allows us to explicitly evaluate the time-ordered integrals of the Dyson series~\eqref{eq:dyson}  (see Appendix~\ref{app:proof}), arriving at
\begin{align}\label{eq:mainresult}
U_I(t)&{=} \sum_{q=0}^{\infty} \sum_z \sum_{{\bf{i}}_q,{\bf{k}}_q} d_z^{({\bf{i}}_q,{\bf{k}}_q)}\e^{-it[x_0,x_1,\ldots,x_{q-1},0]} P_{{\bf{i}}_q}|z\rangle\langle{z}|\nn\\
&{=}\sum_{q=0}^{\infty} \sum_z  \sum_{{\bf{i}}_q} \alpha_z^{({\bf{i}}_q)} P_{{\bf{i}}_q}|z\rangle\langle{z}|\,,
\end{align}
where we have defined  the (time-dependent) coefficients
\begin{align}\label{eq:s}
\alpha_z^{({\bf{i}}_q)}= \sum_{{\bf{k}}_q} d_z^{({\bf{i}}_q,{\bf{k}}_q)}\e^{-it[x_0,x_1,\ldots,x_{q-1},0]}. 
\end{align}
In addition, we have denoted ${{\bf{i}}_q} =(i_1,i_2,\ldots,i_q)$ and ${{\bf{k}}_q} =(k_1,k_2,\ldots,k_q)$ as multi-indices, $P_{{\bf{i}}_q}= P_{i_q} \cdots P_{i_1}$, $d_z^{({\bf{i}}_q,{\bf{k}}_q)}=\prod_{j=1}^q d_{z}^{({\bf{i}}_j,k_j)}$ with $d_{z}^{({\bf{i}}_j,k_j)} = \langle z_{{\bf{i}}_j}|D_{i_j}^{({k_j})}|z_{{\bf{i}}_j}\rangle$ and $\ket{z_{{\bf{i}}_j}}=P_{{\bf{i}}_j}\ket{z}$ for  every $j=0,\ldots,q$ (remembering that the value of $z_{{\bf{i}}_j}$ depends on that of $z$). The inputs for the divided-difference exponential are given by
\begin{align}\label{eq:x}
x_j=E_{z}^{({\bf{i}}_j)}-E_{z}^{({\bf{i}}_q)}-\sum_{\ell=j+1}^{q}\lambda_z^{({\bf{i}}_\ell,k_\ell)}
\end{align}
 for  $j=0,\ldots,q-1$, where $E_{z}^{({\bf{i}}_j)}=\langle z_{{\bf{i}}_j}|H_0|z_{{\bf{i}}_j}\rangle$  and  $\lambda_{z}^{({\bf{i}}_j,k_j)}=\langle z_{{\bf{i}}_j}|\Lambda_{i_j}^{(k_j)}|z_{{\bf{i}}_j}\rangle$  (we have omitted the dependence of $x_0,\ldots,x_{q-1}$ on $z$, ${\bf{i}}_q$ and on ${\bf{k}}_q$ for a lighter notation).  Importantly,  the complex-valued $\e^{-it[x_0,x_1,\ldots,x_{q-1},0]}$ can be calculated efficiently, with computational complexity proportional to $q^2$ (see Ref.~\cite{divDiffCalc} for additional details).
Equation~\eqref{eq:mainresult} is the main technical result of our paper.  The equation is a reformulation of the Dyson series, Eq.~\eqref{eq:dyson}, yet it includes only sums of efficiently-computable terms and is devoid of both integrals and time-ordering operators. 

The current representation of $U_I(t)$ also allows us to obtain an integral-free time-dependent perturbation expansion of the Schr{\"o}dinger-picture time-evolution operator $U(t)$:
\begin{align}\label{eq:U(t)}
U(t){=}e^{-i H_0 t}  U_I (t){=}\sum_{q=0}^{\infty} \sum_z  \sum_{{\bf{i}}_q} \beta_z^{({\bf{i}}_q)}  P_{{\bf{i}}_q}|z\rangle\langle{z}| ,
\end{align}
where 
\begin{align}\label{eq:beta}
\beta_z^{({\bf{i}}_q)}{=} \alpha_z^{({\bf{i}}_q)} e^{-i t E_{z}^{({\bf{i}}_q)}} \!{=}\!\sum_{{\bf{k}}_q}\! d_z^{({\bf{i}}_q,{\bf{k}}_q)}\e^{-it[y_0,y_1,\ldots,y_{q}]},
\end{align}
with $y_j=E_{z}^{({\bf{i}}_j)}-\sum_{\ell=j+1}^{q}\lambda_z^{({\bf{i}}_\ell,k_\ell)}$ for  $j=0,\ldots,q$. 
We arrive at Eq.~\eqref{eq:beta} by using Identity~2 in Appendix~\ref{app:dysonInt}.

Equation~\eqref{eq:U(t)} allows us to explicitly carry out time-dependent perturbation calculations directly in the  Schr{\"o}dinger-picture without the need to invoke the interaction-picture. 
\section{Use cases}
To showcase the immediate applicability of our approach, in what follows we illustrate, by examining a few examples, the utility of Eq.~\eqref{eq:U(t)} in a variety of settings, specifically in scenarios where the time-ordered integrals of the Dyson series are cumbersome to calculate or lead to unnecessary complicated expressions.

{\it Example~1: Transition amplitudes and Fermi's golden rule.}~Using the divided-differences series formulation of the time-evolution operator one may arrive at a rather simple expression for transition amplitudes, $A(z_{\rm in}\to z_{\rm fin},t) =\langle z_{\rm fin}| U(t)|z_{\rm in}\rangle$, between an initial eigenstate $|z_{\rm in}\rangle$ and a final eigenstate $|z_{\rm fin}\rangle$ of the free Hamiltonian $H_0$ due to the perturbation effect of $V(t)$, cf. Eq.~\eqref{eq:ham}. Using Eq.~\eqref{eq:U(t)} we find
\beq
A(z_{\rm in}\to z_{\rm fin},t) =\sum_{q=0}^\infty \!\!\!\!\!\!\!\!\!\!\!\!\!\!\!\!\!\!\!\!\!\!\!\!\!\!\!\!\!\! \sum_{\;\;\;\;\;\;\;\;\;\;\;\;\;\;\;\;\;\;\;\;\;\;\; {\bf{i}}_q:\langle z_{\rm fin} | P_{{\bf{i}}_q} |z_{\rm in}\rangle=1}
\!\!\!\!\!\!\!\!\!\!\!\!\!\!\!\!\!\!\!\!\!\!\!\!\!\!\!\!\!\!
\beta^{({\bf{i}}_q)}_{z_{\rm in}} 
\,,
\eeq
where the second sum is over all multi-indices ${\bf{i}}_q$ that obey $ P_{{\bf{i}}_q}\ket{z_{\rm in}}=|z_{\rm fin}\rangle$.

Consider, for example, for the canonical, yet non trivial, case of the non-Hermitian Hamiltonian $H=H_0 + \gamma F \e^{i E t}$ where $\gamma$ is a  perturbation parameter and $F$ a permutation matrix such that $F|z_{\rm in}\rangle=|z_{\rm fin}\rangle$ and $F^2=\mathbb{1}$. While the standard Dyson-series-based derivation of the transition amplitude beyond  the first order expansion $q=1$ (that is Fermi's golden rule) is cumbersome to derive  and to calculate (see, e.g., Refs.~\cite{Landau,schiffBook}), the present formulation allows us to immediately write a succinct expression for transition amplitudes that are accurate to {\it any} order in $\gamma$. In particular, the $q$-th order contribution for the transition amplitude is given very neatly by 
\begin{align}
A^{(q)}(z_{\rm in} \to z_{\rm fin},t) =
\gamma^q \e^{-it E_{z_{\rm in}}} \e^{-it[ qE, (q-1) E,\ldots,E,0]},
\end{align}
for odd values of $q$,  and can be calculated with $O(q^2)$ basic operations (for even values of $q$, $A^{(q)}(z_{\rm in} \to z_{\rm fin},t) =0$). The case $q=1$ corresponds to Fermi's golden rule~\cite{Merzbacher}. This simple expression showcases how the new formulation of the Dyson series can be used to explore properties of time-varying Hamiltonian systems that arise from perturbation orders that have not explored with prior methods. 

\begin{figure}[t!]
\begin{center}
\includegraphics[width=1\columnwidth]{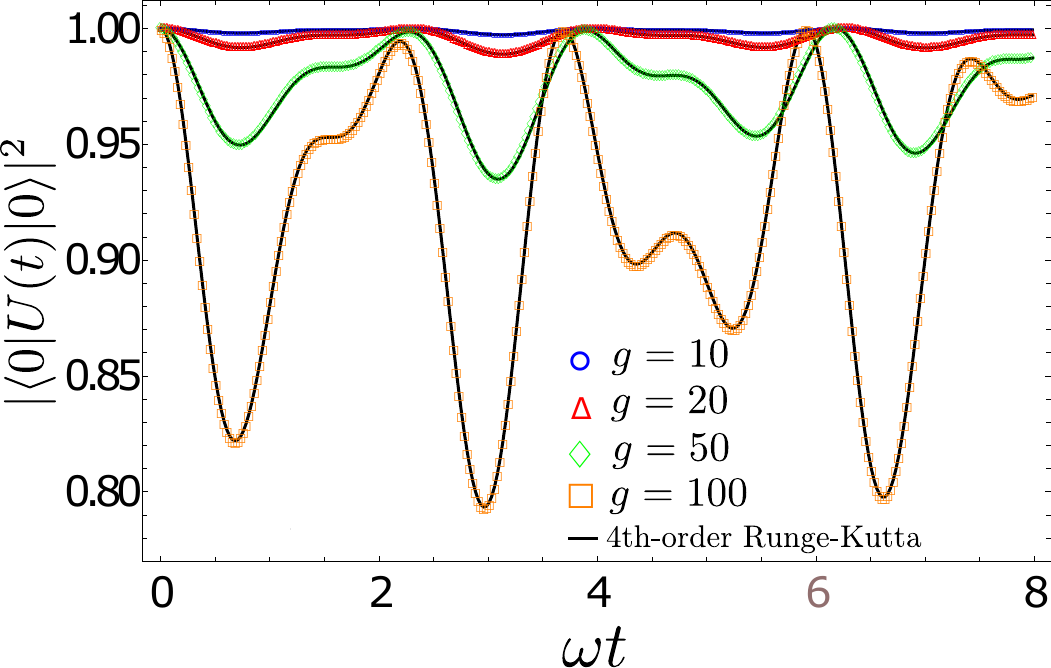}
\caption{The probability to remain in the initial state $|0\rangle$ (eigenstate of $\sigma_z$ with eigenvalue $+1$) for a highly oscillating (here, ${\omega_0=2 \omega=200}$) two-level system for different values of time $t$. The figure shows numerically-exact calculations using the proposed method for a wide range of coupling strengths. The black lines passing through the data points correspond to numerically-exact solutions obtained using 4th-order Runge-Kutta simulations carried out independently to verify the correctness of the our results.}
\label{fig:2l}
\end{center}
\end{figure}

{\it Example~2: Time-oscillating two-level Hamiltonian system.}~To further illustrate the practical power of the integral-free description of the Dyson series and its ability to assist in uncovering new physics, we next consider the dynamics of two-level systems driven by highly oscillating time-dependent Hamiltonians of the form \hbox{$H=\omega_0 \sigma_z + g \sigma_x \cos \omega t$}.  Despite its apparent simplicity, calculating dynamical properties of this system still poses challenges to numerical integration methods whenever the driving oscillations are very fast, and it is an open problem that continues to be a very active area of research, for example in the study of evolution of qubits undergoing
quantum gates~\cite{giscardDynamics} or in Bloch-Siegert systems, which play an important role in atomic physics~\cite{Frasca} (It should be noted that in certain cases two-level systems admit analytical solutions in which case a numerical series expansion is not strictly necessary, see e.g.~\cite{SchmidtFloquet}).  Our formulation on the other hand results in a compact description of the unitary operator (more information is given in Appendix~\ref{app:timeoc2}, which in turn allows for a numerically-exact calculation of various properties of the system, such as transition amplitudes, even in regions of large $\omega$ where direct integration of the Sch{\"o}rdinger equation becomes challenging. An example is given in Fig.~\ref{fig:2l} wherein a few numerically-exact results are given for various scenarios, as calculated by summing over the appropriate divided-difference terms.  The figure also depicts independently derived numerically-exact solutions obtained using 4th-order Runge-Kutta simulations~\cite{RK} carried out independently. As is evident, there is an excellent agreement between the two methods.

{\it Example~3: Time-oscillating infinite-dimensional Hamiltonian system.}~Next, we work out the expansion of the time-evolution operator for a particle moving under the influence of a harmonic potential perturbed by a periodic time-dependent anharmonic term, namely, 
\beq
H(t)=\frac{\hat{p}^2}{2m} + \frac{1}{2} m \omega^2 \hat{x}^2 + \frac1{4} \Gamma \cos (\Omega t) ~\hat{x}^4 \,.
\eeq
Identifying the static component as $H_0=\hat{p}^2/2m  + m \omega^2 \hat{x}^2/2$, we rewrite the Hamiltonian in terms of the harmonic annihilation-creation operators (recall that $\hbar=1$) as
\beq\label{eq:Hanhar}
H(t)= \omega \big(\hat{a}^\dagger \hat{a}+\frac1{2}\big) + \underbrace{\frac1{4} \Gamma \cos (\Omega t) \left(\frac{1}{2 m\omega}\right)^2(\hat{a}^\dagger + \hat{a})^4}_{V(t)} \,.
\eeq
To express $V(t)$ in terms of generalized permutation operators, using the harmonic oscillator eigenstates $\{\ket{n}\}$ as the computational basis states, we rewrite the anharmonic operator $\left(\hat{a}^\dagger + \hat{a}\right)^4$ as
\hbox{$\left(\hat{a}^\dagger + \hat{a}\right)^4 = \sum_{i \in \{0,\pm 2,\pm 4\}} D_i P_i$} where \hbox{$P_i= \sum_n |n+i\rangle\langle n|$} and the matrix elements of the diagonal operators $D_i$ are given by
\bea
{(D_{-4})}_{nn} &=&\sqrt{n(n-1)(n-2)(n-3)}  \,,\\\nonumber
{(D_{-2})}_{nn}&=& \sqrt{n(n-1)}(4n-2) \,,\\\nonumber
{(D_0)}_{nn} &=& 3(2n^2+2n+1)  \,,\\\nonumber
{(D_{2})}_{nn}&=& \sqrt{(n+1)(n+2)}(4n+6) \,,\\\nonumber
{(D_{4})}_{nn}&=& \sqrt{(n+1)(n+2)(n+3)(n+4)}  \,.\\\nonumber
\eea
In terms of the above operators, the Hamiltonian can be written as
\beq
H(t)= H_0+ \gamma \sum_{i}\sum_{k=\pm 1}  \e^{-i k \Omega t} D_i P_i
\eeq
where $H_0= \omega \sum_n (n+\frac1{2})  |n\rangle \langle n|$ and  $\gamma=\frac1{8} \Gamma \left(\frac{1}{2 m\omega}\right)^2$. 
Having cast the Hamiltonian in the appropriate form, it is straightforward to calculate the $\beta^{({\bf{i}}_q)}_{n}$ terms in the divided-differences expansion of $U(t)$, as per Eq.~(\ref{eq:beta}). For example for $q=1$ we obtain
\begin{align}
\beta^{(i_1)}_{n}&=\gamma \e^{- i t E_{n+i_1}}\\\nn &\times (D_{i_1})_{nn} (\e^{- i t [\Omega-i_1 \omega,0]}+ \e^{- i t [-(\Omega+i_1\omega),0]})
\end{align}
where $i_1 \in \{0,\pm 2,\pm 4\}$.
Our formulation provides an easy way to compute the state of the system after time $t$. For an initial state $\ket{\psi(0)}=|n\rangle$, we get the analytic closed-form expression $|\psi(t)\rangle=U(t)\ket{n}=\sum_{{\bf{i}}_q} \beta^{({\bf{i}}_q)}_{n} P_{{\bf{i}}_q}|n\rangle$ (see Appendix~\ref{app:timeocInf} for additional information and further analysis).

\section{Summary and discussion}
To conclude, in this work, we devised an alternative approach to time-dependent perturbation theory in quantum mechanics that allows one to readily calculate expansion terms. 
We derived an expansion that is equivalent to the usual Dyson series but which includes only sums of closed-form analytical simple-to-calculate expressions rather than the usual Dyson multi-dimensional time-ordered integrals. The terms at every expansion order in our new formulation coincide with those of the standard Dyson series, except that the Dyson integrals at every order replaced with finite sums. Therefore, both series share the same convergence criteria~\cite{dsd1,dsd2,dsd3,dsd4}.  However, our new formalism allowed us to write an integral-free perturbation expansion for the time-evolution operator. We illustrated the utility of our approach by working out a number of use cases and calculated the series coefficients for a number of examples for which the usual Dyson series calculation is cumbersome, demonstrating the functionality and practicality of our approach. 

Another area in which the divided-differences expansion can be applied, which we have not explored here, is quantum algorithms -- algorithms designed to be executed on quantum computers. Specifically, the divided-differences expansion was recently shown to be a valuable tool in the derivation of quantum algorithm devised to simulate the time-evolution of quantum states evolving under time-independent~\cite{2020arXiv200602539K} and  time-dependent~\cite{Chen_2021} Hamiltonians.
There, it was shown that the divided-differences expansion allows for the time- evolution operator to be written as a sum of generalized permutation matrices, equivalently a linear combination of unitary (LCU) operators. As such, this representation of the time-evolution operator lends itself naturally to the quantum LCU lemma~\cite{Berry1} which provides a prescription for efficiently simulating such operators on quantum circuits. We leave that for future work. 

We believe that the formulation introduced here will prove itself to be a powerful tool in the study of time-dependent quantum many-body systems.

\begin{acknowledgments}
This work is supported by the U.S. Department of Energy (DOE), Office of Science, Basic Energy Sciences (BES) under Award No. DE-SC0020280. 
\end{acknowledgments}

\bibliography{refs}

\appendix
\begin{widetext} 
\section{Derivation of an integral-free form for the Dyson series}\label{app:proof}
Here, we derive the expression
\begin{align}\label{eq:mainresultApp}
U_I(t)&{=} \sum_{q=0}^{\infty} \sum_z \sum_{{\bf{i}}_q,{\bf{k}}_q} d_z^{({\bf{i}}_q,{\bf{k}}_q)}\e^{-it[x_0,x_1,\ldots,x_{q-1},0]} P_{{\bf{i}}_q}|z\rangle\langle{z}|.
\end{align}
Using the form of $H_I(t)$
\begin{align}\label{eq:hint3App}
H_I(t)=\sum_{i=1}^M \sum_{k=1}^{K}  \e^{i H_0^{(i,k)} t} D_i^{(k)} P_i \e^{-i H_0 t} \,,
\end{align}
 we can write the Dyson series for $U_I(t)$
as 
\begin{align}
U_I(t)&=\sum_{q=0}^{\infty} (-i)^q \int_0^{t} \rmd t_q \cdots \int_0^{t_2} \rmd t_1 \left( \sum_{i_q=1}^M\sum_{k_q=1}^K \e^{i H_0^{(i_q,k_q)} t_q} D_{i_q}^{({k_q})} P_{i_q} \e^{-i H_0 t_q}\right)\cdots\\\nonumber & 
\;\;\;\;\;\;\;\;\;\;\;\;\;\;\;\;\;\;\;\;\;\;\;\;\;\;\;\;\;\;\;\;\;\;\;\;\;\;\;\;
\;\;\;\;\;\;\;\;\;\;\;\;\;\;\;\;\;\;\;\;\;\;\;\;\;\;\;\;\;\;\;\;\;\;\;\;\;\;\;\;
\;\;\;\;\;
\cdots \left( \sum_{i_1=1}^M\sum_{k_1=1}^K \e^{i H_0^{(i_1,k_1)} t_1} D_{i_1}^{({k_1})} P_{i_1} \e^{-i H_0 t_1}\right)  
\\\nonumber
&=\sum_{q=0}^{\infty} \sum_{{\bf{i}}_q}\sum_{{\bf{k}}_q} (-i)^q \int_0^{t} \rmd t_q  \cdots \int_0^{t_2}\rmd t_1 \left( \e^{i H_0^{(i_q,k_q)} t_q} D_{i_q}^{({k_q})} P_{i_q} \e^{-i H_0 t_q}  \cdots  \e^{i H_0^{(i_1,k_1)} t_1} D_{i_1}^{({k_1})} P_{i_1} \e^{-i H_0 t_1}\right),
\end{align}
where ${{\bf{i}}_q} =(i_1,i_2,\ldots,i_q)$ and ${{\bf{k}}_q} =(k_1,k_2,\ldots,k_q)$ are multi-indices.  Acting with $U_I(t)$ on an arbitrary computational basis state $|z\rangle$, we get
\begin{align}
U_I(t)|z\rangle&=\sum_{q=0}^{\infty} \sum_{{\bf{i}}_q}\sum_{{\bf{k}}_q} (-i)^q \int_0^{t} \rmd t_q  \cdots \int_0^{t_2}\rmd t_1\e^{i E_{z}^{({\bf{i}}_{q},k_q)} t_q}  d_{z}^{({\bf{i}}_q,k_q)} \e^{-i E_{z}^{({\bf{i}}_{q-1})} t_q}  \cdots \e^{i E_{z}^{({\bf{i}}_1,k_1)}t_1} d_{z}^{({\bf{i}}_1,k_1)} \e^{-i E_{z} t_1} P_{{\bf{i}}_q}|z\rangle
\nonumber\\
&=\sum_{q=0}^{\infty} \sum_{{\bf{i}}_q}\sum_{{\bf{k}}_q} d_z^{({\bf{i}}_q,{\bf{k}}_q)} \Bigl((-i)^q \int_0^{t} \rmd t_q  \cdots \int_0^{t_2}\rmd t_1\e^{-i (E_{z}^{({\bf{i}}_{q-1})}-E_{z}^{({\bf{i}}_{q},k_q)}) t_q}  \cdots \e^{-i (E_{z}-E_{z}^{({\bf{i}}_1,k_1)}) t_1}\Bigr) P_{{\bf{i}}_q}|z\rangle\label{eq:U_I1App}\,,
\end{align}
where we have defined $P_{{\bf{i}}_q}= P_{i_q} \cdots P_{i_1}$ and $\ket{z_{{\bf{i}}_j}}=P_{{\bf{i}}_j}\ket{z}$ for  every $j=0,\ldots,q$ (remembering that $z_{{\bf{i}}_j}$ depends on $z$). Moreover, we define  $E_{z}^{({\bf{i}}_j)}=\langle z_{{\bf{i}}_j}|H_0|z_{{\bf{i}}_j}\rangle$  and similarly
\beq
E_{z}^{({\bf{i}}_j,k_j)}=\langle z_{{\bf{i}}_j}|H_0^{(i_{j},k_j)}|z_{{\bf{i}}_j}\rangle=
E_{z}^{({\bf{i}}_j)} + \lambda_{z}^{({\bf{i}}_j,k_j)}\,,
\eeq 
with $\lambda_{z}^{({\bf{i}}_j,k_j)}=\langle z_{{\bf{i}}_j}|\Lambda_{i_j}^{(k_j)}|z_{{\bf{i}}_j}\rangle$. 
In addition, we have denoted 
$d_{z}^{({\bf{i}}_j,k_j)} = \langle z_{{\bf{i}}_j}|D_{i_j}^{({k_j})}|z_{{\bf{i}}_j}\rangle$ and \hbox{$d_z^{({\bf{i}}_q,{\bf{k}}_q)}=\prod_{j=1}^q d_{z}^{({\bf{i}}_j,k_j)}$}. 
Following Identity~1 in Appendix~\ref{app:dysonInt}, the  integrals appearing inside the parentheses in  Eq.~\eqref{eq:U_I1App} evaluate explicitly to the divided-differences exponential, that is,
\begin{align}\label{eq:explicitApp}
(-i)^q \int_0^{t} \rmd t_q  \cdots \int_0^{t_2}\rmd t_1\e^{-i (E_{z}^{({\bf{i}}_{q-1})}-E_{z}^{({\bf{i}}_{q},k_q)}) t_q}  \cdots \e^{-i (E_{z}-E_{z}^{({\bf{i}}_1,k_1)}) t_1}=\e^{-it[x_0,x_1,\ldots,x_{q-1},0]} \,,
\end{align}
with
\begin{align}\label{eq:xApp}
x_j=\sum_{\ell=j+1}^{q} E_{z}^{({\bf{i}}_{\ell-1})}-E_{z}^{({\bf{i}}_{\ell},k_\ell)}=E_{z}^{({\bf{i}}_j)}-E_{z}^{({\bf{i}}_q)}-\sum_{\ell=j+1}^{q}\lambda_z^{({\bf{i}}_\ell,k_\ell)} \;\;\text{for}\;\; j=0,\ldots,q-1
\end{align}
(we have omitted the dependence of $x_0,\ldots,x_{q-1}$ on $z$, ${\bf{i}}_q$ and on ${\bf{k}}_q$ for a lighter notation).    Using Eq.~\eqref{eq:explicitApp}, we can write the action of $U_I(t)$ on an arbitrary computational basis state in an integral-free form as
\beq
U_I(t)|z\rangle = \sum_{q=0}^{\infty} \sum_{{\bf{i}}_q}\sum_{{\bf{k}}_q} d^{({\bf{i}}_q,{\bf{k}}_q)}_z\e^{-it[x_0,x_1,\ldots,x_{q-1},0]} P_{{\bf{i}}_q}|z\rangle\,,
\eeq
with the inputs to the divided-differences exponential given in Eq.~\eqref{eq:xApp}. We thus arrive at:
\beq
U_I(t)= \sum_{q=0}^{\infty} \sum_z  \sum_{{\bf{i}}_q} \sum_{{\bf{k}}_q} d_z^{({\bf{i}}_q,{\bf{k}}_q)}\e^{-it[x_0,x_1,\ldots,x_{q-1},0]} P_{{\bf{i}}_q}|z\rangle\langle{z}|,
\eeq
as claimed.

\section{Proofs of two identities of the exponent of divided-differences\label{app:dysonInt}}

\noindent{\bf Identity 1:}~
\begin{align}\label{eq:eye1App}
(-i)^q\int_0^{t} \rmd t_q \cdots \int_0^{t_2}\rmd t_1   \e^{-i ( \gamma_q t_q+ \cdots   + \gamma_1 t_1)}=\e^{-i t [x_0,x_1,\ldots,x_{q-1},0]}
\end{align}
where  $x_j=\sum_{k=j+1}^{q}\gamma_k$. This identity is a variant of what is often known as the Hermite-Gennochi formula~\cite{deboor:05}. \vspace{0.2cm}\\
\noindent {\bf Proof}:
We start with the left-hand-side of Eq.~\eqref{eq:eye1App}. Making a change of variables $t_j = t s_j$ we get
\begin{align}\label{eq:sApp}
(-i)^q\int_0^{t} \rmd t_q \cdots \int_0^{t_2}\rmd t_1   \e^{-i ( \gamma_q t_q+ \cdots   + \gamma_1 t_1)}&=(-it)^q\int_0^{1} \rmd s_q \cdots \int_0^{s_2}\rmd s_1   \e^{-it ( \gamma_q s_q+ \cdots   + \gamma_1 s_1)}.
\end{align}
Next we prove by induction  that 
\begin{align}\label{eq:dysApp}
\e^{-i t [x_0,x_1,\ldots,x_{q-1},0]}=(-it)^q\int_0^{1} \rmd s_q \cdots \int_0^{s_2}\rmd s_1   \e^{-it ( \gamma_q s_q+ \cdots   + \gamma_1 s_1)}
\end{align}
where  $x_j=\sum_{k=j+1}^{q}\gamma_k$.\vspace{0.3cm}\\
\noindent{\it Base step (proving for $q=1$)}: Starting from the left-hand-side of Eq.~\eqref{eq:dysApp} we have
\begin{align}\label{eq:indq1App}
(-it)\int_{0}^{1}\rmd s_1 \e^{-it\gamma_{1}s_1}&=\frac{(-it)}{\gamma_{1}}\int_{0}^{\gamma_{1}}\rmd\xi \e^{-it\xi}=\frac{1}{\gamma_{1}}\int_{0}^{\gamma_{1}}\rmd\xi \frac{\rmd}{\rmd\xi}\e^{-it\xi}=\frac{\e^{-it \gamma_{1}}-1}{\gamma_{1}}=\e^{-it [0,\gamma_{1}]}=\e^{-it [\gamma_{1},0]}
\end{align}
as required. In the first equality of Eq.~\eqref{eq:indq1App}, we changed the integration variable $\xi=s_1\gamma_{1}$, so that $\xi$ ranges from $0$ to $\gamma_1$, and $\gamma_{1}\rmd s_1=\rmd\xi $.  \\
\noindent{\it Hypothesis step}: Next, we assume the validity of Eq.~\eqref{eq:dysApp}.\\
\noindent{\it Induction step}: Based on this assumption, we now prove that
\begin{align}
(-it)^{q+1}\int_0^{1} \rmd s_{q+1} \cdots \int_0^{s_2}\rmd s_1   \e^{-it ( \gamma_{q+1} s_{q+1}+ \cdots   + \gamma_1 s_1)}=\e^{-i t [x_0,x_1,\ldots,x_{q},0]}
\end{align}
with $x_j= \sum_{k=j+1}^{q+1} \gamma_k$. Changing the integration variable $s_1$ to $\xi=\gamma_{q+1} s_{q+1}+ \cdots   + \gamma_1 s_1$, so that $\xi$ ranges from $\gamma_{q+1}  s_{q+1}+ \cdots   + \gamma_{2}  s_2$ to $\gamma_{q+1}  s_{q+1}+ \cdots   + (\gamma_2+\gamma_{1}) s_2$, and $\gamma_{1} \rmd s_1=\rmd\xi $ we get
\begin{align}
&(-it)^{q+1}\int_0^{1} \rmd s_{q+1} \cdots \int_0^{s_2}\rmd s_1   \e^{-it \big(\gamma_{q+1}  s_{q+1}+ \cdots   + \gamma_{1}  s_1\big)}
\nonumber\\&=(-it)^{q+1}\frac1{\gamma_{1} }\int_0^{1} \rmd s_{q+1} \cdots \int_{\gamma_{q+1}  s_{q+1}+ \cdots   + \gamma_{2}  s_2}^{\gamma_{q+1}  s_{q+1}+ \cdots   + (\gamma_2+\gamma_{1}) s_2}\rmd \xi \e^{-it \xi}
\nonumber\\&=(-it)^{q}\frac1{\gamma_{1}}\int_0^{1} \rmd s_{q+1} \cdots \int_{\gamma_{q+1}  s_{q+1}+ \cdots   + \gamma_{2}  s_2}^{\gamma_{q+1}  s_{q+1}+ \cdots   + (\gamma_2+\gamma_{1}) s_2}\rmd \xi \frac{\rmd}{\rmd\xi}\e^{-it \xi}
\nonumber\\&=(-it)^{q}\frac1{\gamma_{1}}\int_0^{1} \rmd s_{q+1} \cdots \int_0^{s_3}\rmd s_2  \Big(\e^{-it \big(\gamma_{q+1}  s_{q+1}+ \cdots   + (\gamma_2+\gamma_{1}) s_2\big)}
-\e^{-it \big(\gamma_{q+1}  s_{q+1}+ \cdots   + \gamma_{2}  s_2\big)}\Big)
\nonumber\\&=\frac1{\gamma_{1}}\Bigg(\underbrace{(-it)^{q}\int_0^{1} \rmd s_{q+1} \cdots \int_0^{s_3}\rmd s_2  \e^{-it \big(\gamma_{q+1}  s_{q+1}+ \cdots   + (\gamma_2+\gamma_{1}) s_2\big)}}_{(*)}
\nonumber\\&\qquad\;\;\;
-\underbrace{(-it)^{q}\int_0^{1} \rmd s_{q+1} \cdots \int_0^{s_3}\rmd s_2  \e^{-it \big(\gamma_{q+1}  s_{q+1}+ \cdots   + \gamma_{2}  s_2\big)}}_{(**)}\Bigg),
\end{align}
where from the induction assumption,
\begin{align}
(*)&=\e^{-i t [x_0, x_2,\ldots,x_q,0]},\;\;\text{and}
\;\;
(**)=\e^{-i t [ x_1,x_2, \ldots,x_q,0]},
\end{align}
with $x_j=\sum_{k=j+1}^{q+1}\gamma_k$. Therefore, we obtain,
\begin{align}
&(-it)^{q+1}\int_0^{1} \rmd s_{q+1} \cdots \int_0^{s_2}\rmd s_1   \e^{-it \big(\gamma_{q+1}  s_{q+1}+ \cdots   + \gamma_{1}  s_1\big)}
\nonumber\\&=\frac{\e^{-i t [x_0, x_2,\ldots,x_q,0]}-\e^{-i t [ x_1,x_2, \ldots,x_q,0]}}{\gamma_{1} }=\frac{\e^{-i t [x_2,\ldots,x_q,0,x_0]}-\e^{-i t [ x_2, \ldots,x_q,0,x_1]}}{x_0-x_1 }
\nonumber\\&=  \e^{-i t [x_2,\ldots,x_q,0,x_0,x_1]}=\e^{-i t [x_0,x_1,x_2,\ldots,x_q,0]}.
\end{align}
This concludes the proof.\vspace{0.3cm}\\
\noindent {\bf Identity 2:}~Given an arbitrary multi-set of inputs $\{x_0,\ldots,x_q\}$, 
\beq\label{eq:deltaApp}
\e^{- i t [x_0,\ldots,x_q]} =\e^{- i t x} \e^{- i t [\Delta_0,\ldots,\Delta_q]} \,,
\eeq
where $x$ is an arbitrary constant and $\Delta_j = x_j -x$. \vspace{0.2cm}\\
\noindent {\bf Proof}: We prove Eq.~\eqref{eq:deltaApp}. By definition~\cite{deboor:05}, 
\beq
\e^{- i t [x_0,\ldots,x_q]} = \sum_j \frac{\e^{-i t x_j}}{\prod_{ k \neq j} (x_j -x_k)} \,
\eeq
(assuming for now that all inputs are distinct). It follows then that
\begin{align}
\e^{- i t [x_0,\ldots,x_q]}= \sum_j \frac{\e^{-i t (\Delta_j+x)}}{\prod_{ k \neq j} (\Delta_j -\Delta_k)}  =\e^{-i t x} \sum_j \frac{\e^{-i t \Delta_j}}{\prod_{ k \neq j} (\Delta_j -\Delta_k)}=\e^{-i t x} \e^{- i t [\Delta_0,\ldots,\Delta_q]} \,.
\end{align}
This result holds for arbitrarily close inputs and can be easily generalized to the case where inputs have repeated values, as claimed.

\section{Time-oscillating two-level Hamiltonian systems\label{app:timeoc2}}
Consider a time-dependent qubit system whose Hamiltonian has the form
$H=\omega_0 \sigma_z + g \sigma_x \cos \omega t$. We identify $H_0= \omega_0 \sigma_z $ (hence the computational basis is the Pauli-Z eigenbasis $\{\ket{z}:z=0,1\}$) and \hbox{$V(t)=g \sigma_x \cos \omega t$}. The time-dependent component $V(t)$ contains a one permutation operators ($M=1$), $P_1=\sigma_x$, whose associated diagonal operators, $D_1=g \cos \omega t\,\mathbb{1}=(g/2) (e^{i\omega t}+e^{-i\omega t})\mathbb{1}$ has two exponential terms ($K=2$) such that $D_1=e^{i\omega t}D_1^{(1)}+e^{-i\omega t}D_1^{(-1)}$ with $D_1^{(1)}=D_1^{(-1)}=(g/2)\mathbb{1}$. For reasons that would be come clearer later we index the latter operators by $k=\pm1$ (instead of $k=1,2$). 

To write the time evolution operator, we first note that since there is only one off-diagonal operator in the Hamiltonian, there is also one sequence of off-diagonal operators per expansion order  $q$, explicitly:  
\beq
P_{{\bf{i}}_q}= \sigma_x^{q}=
\begin{cases}
\sigma_x,\;\;\text{for odd $q$}\\
\mathbb{1},\;\;\text{for even $q$}
\end{cases}
\eeq
Moreover $\beta_z^{(q)}$, defined in Eq.~(12) in the main text, is given by
\beq
\beta_z^{(q)}=\Big(\frac{g}{2}\Big)^q \sum_{{\bf k}_q}e^{-it [y_0,\ldots,y_q]} ,
\eeq
with $y_j=E_{z}^{(j)}-\omega\Sigma{\bf{k}}_{j+1:q}$ for  $j=0,\ldots,q$, where $E_{z}^{(j)}=\bra{z}\sigma_x^{j} D_0 \sigma_x^{j}\ket{z}=(-1)^{z+j}\omega_0$ and $\Sigma{\bf{k}}_{j+1:q}$ is a shorthand notation to $\sum_{\ell=j+1}^q k_{\ell}$. The last two equations define the time evolution operator of the system for any order expansion $q$, specifically,
\beq\label{U(t) driven qubit}
U(t)=\sum_{z=0,1}\Big(\sum_{q=0}^\infty\beta_z^{(2q)}(t)\Big) \ket{z}\bra{z}+\sum_{z=0,1}\Big(\sum_{q=0}^\infty\beta_z^{(2q+1)}(t)\Big)  \ket{\bar{z}}\bra{z} \,
\eeq
where $|\bar{z}\rangle = |1-z\rangle$. 

\section{Time-oscillating infinite-dimensional Hamiltonian system\label{app:timeocInf}}
Below we provide additional calculations for the time-dependent anharmonic oscillator
\beq\label{eq:HanharApp}
H(t)= \omega \left(\hat{a}^\dagger \hat{a}+\frac1{2}\right) + \frac1{4} \Gamma \cos (\Omega t) \left(\frac{1}{2 m\omega}\right)^2(\hat{a}^\dagger + \hat{a})^4
\eeq
discussed in the main text. 
Having computed the coefficients, $\beta^{({\bf{i}}_q)}_{n}$ (see Eq.~(19) in the main text), we may plot, as we do in  Fig.~\ref{fig:anhar}(a), the population of every mode at various times. In the figure, the populations are given at $t=0.04$ for $\omega=1, \Omega=2, \gamma=0.02$ with the initial state $|n=4\rangle$ and an expansion cutoff of $Q=5$.
To ascertain the accuracy of the Dyson series truncated at different cutoff orders $Q$, we may also contrast the divided-differences expansion with exact-numerical results obtained via direct integration of the Schr{\"o}dinger equation. In Fig.~\ref{fig:anhar}(b) we plot the infidelity $1-|\langle \psi(t)|\psi_Q(t)\rangle|^2$ between the state $|\psi(t)\rangle$, the solution of the time-dependent Schr{\"o}dinger equation of this system, and the state as obtained by using divided-differences expansion for $U(t)$, $\ket{\psi_Q(t)}=U_Q(t)\ket{4}$, that is, the state obtained by evolving $\ket{4}$ under $U(t)$ given by
\begin{align}\label{eq:U(t)App}
U(t){=}e^{-i H_0 t}  U_I (t){=}\sum_{q=0}^{\infty} \sum_z  \sum_{{\bf{i}}_q} \beta_z^{({\bf{i}}_q)}  P_{{\bf{i}}_q}|z\rangle\langle{z}| \,,
\end{align}
as a function of evolution time $t$. 
\begin{figure}[htp]
\begin{center}
\includegraphics[width=0.45\textwidth]{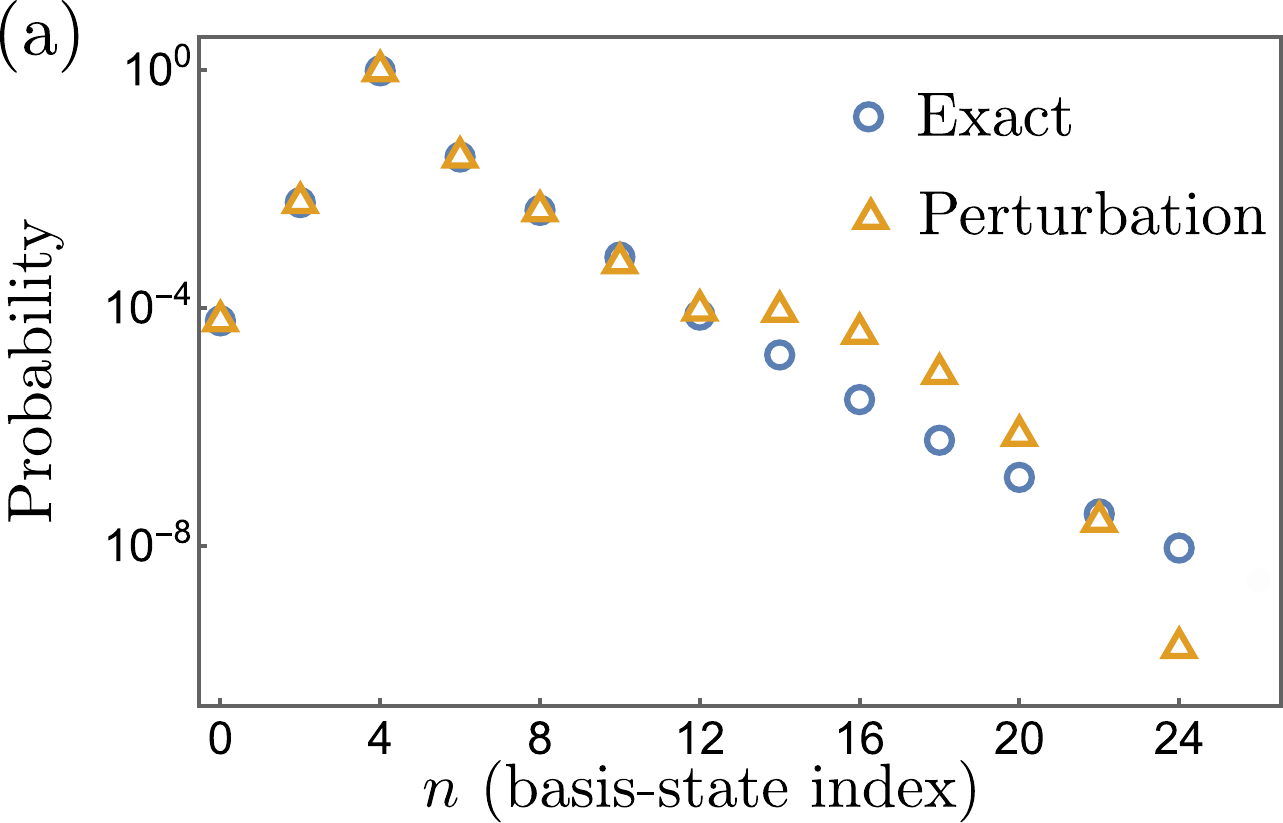}\hspace{0.3cm}
\includegraphics[width=0.47\textwidth]{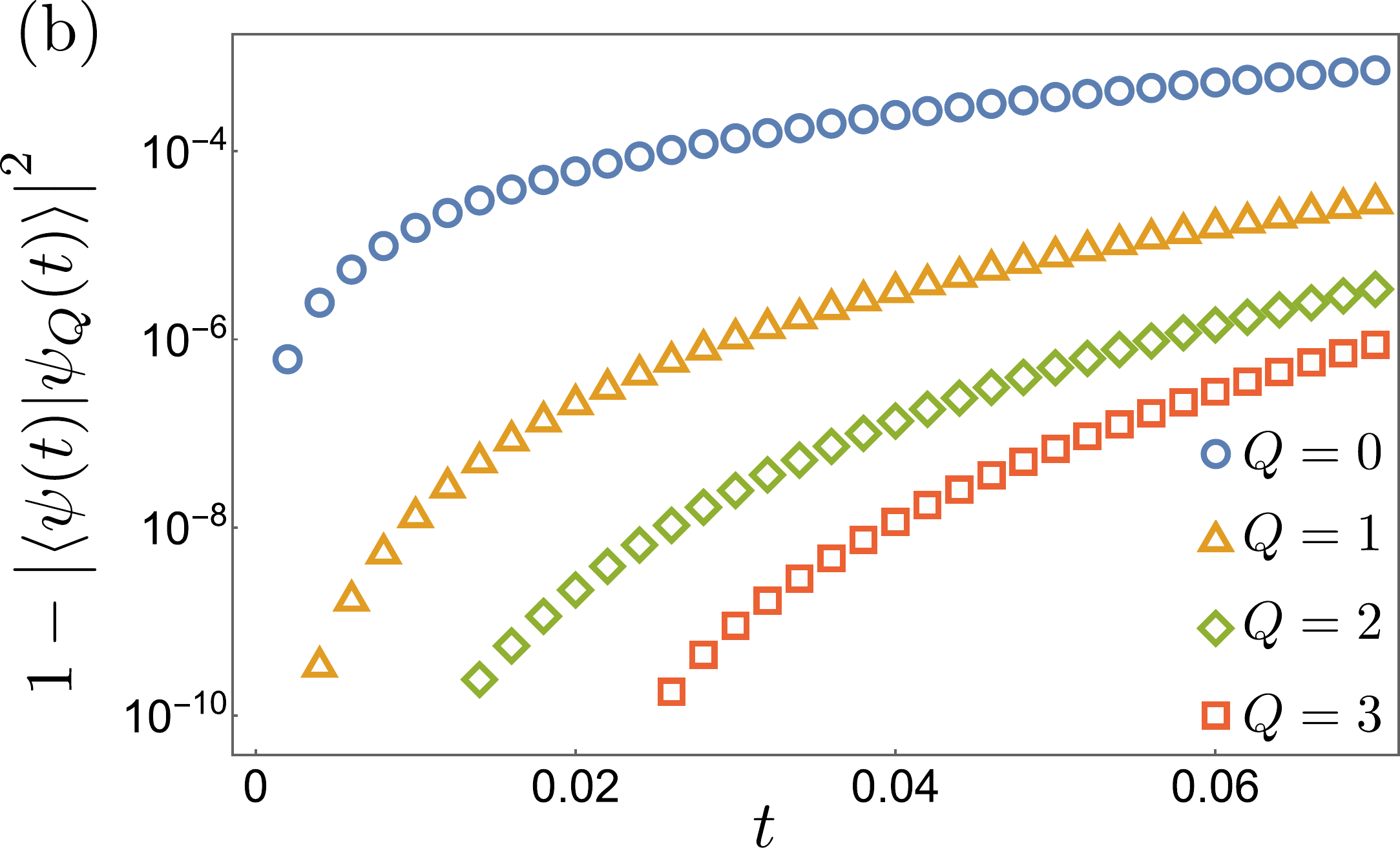}
\caption{(a) Probabilities of the various modes at $t=0.04$ for the anharmonic Hamiltonian, Eq.~(\ref{eq:HanharApp}). Here, $\omega=1, \Omega=2, \gamma=0.02$ and the initial state is $|n=4\rangle$. The blue circles are obtained from numerical integration of the Schr{\"o}dinger equation while the orange triangles correspond to the divided-differences expansion up to order $Q=5$. (b) Infidelity $1-|\langle \psi(t)|\psi_Q(t)\rangle|^2$ as a function of time  The various curves correspond to different truncation orders $Q=0,\ldots,3$. As we expect, the higher the expansion order is, the better the approximation becomes.}
\label{fig:anhar}
\end{center}
\end{figure}
\end{widetext}

\end{document}